\documentclass[journal,10pt,twocolumn]{IEEETran}

\usepackage{threeparttable}
\usepackage{cite}
\usepackage{amsmath,amssymb,amsfonts}
\usepackage{algorithmic}
\usepackage{graphicx}
\usepackage{textcomp}
\usepackage{indentfirst}
\usepackage{csquotes}
\usepackage{multirow}
\usepackage{tabularx}
\usepackage{float}
\usepackage{color}
\usepackage{comment}
\usepackage{caption2}
\usepackage{algorithm2e}
\usepackage{amsmath}
\usepackage{physics}
\usepackage{comment}
\usepackage{dblfloatfix}

\usepackage{tikz}
\newcommand*\circledwhite[1]{\tikz[baseline=(char.base)]{
            \node[shape=circle,draw,inner sep=0.05pt] (char) {\textcolor{black}{#1}};}}

\providecommand{\todo}[1]{{\protect\color{red}\noindent {\bf [TODO]}\emph{#1} {\bf [/TODO]}}}

\newcommand{\mySimulatorName}{SANQ}

\begin{document}

\title{\huge\mySimulatorName: A \underline{S}imulation Framework for \underline{A}rchitecting \underline{N}oisy Intermediate-Scale \underline{Q}uantum Computing System}

\author{Gushu Li, Yufei Ding, and Yuan Xie \\
Unversity of California, Santa Barbara, CA, 93106, USA   \\  
\{gushuli,yufeiding,yuanxie\}@ucsb.edu}

\maketitle

\begin{abstract}

To bridge the gap between limited hardware access and the huge demand for experiments for Noisy Intermediate-Scale Quantum~(NISQ) computing system study,
a simulator which can capture the modeling of both the quantum processor and its classical control system to realize early-stage evaluation and design space exploration, is naturally invoked but still missing. 
This paper presents \mySimulatorName, a \textbf{S}imulation framework for \textbf{A}rchitecting \textbf{N}IS\textbf{Q} computing system. 
\mySimulatorName~consists of two components, 1) an optimized noisy quantum computing~(QC) simulator with flexible error modeling accelerated by eliminating redundant computation, and 2) an architectural simulation infrastructure to construct behavior models for evaluating the control systems. 
\mySimulatorName~is validated with existing NISQ quantum processor and control systems to ensure simulation accuracy. 
It can capture the variance on the QC device and simulate the timing behavior precisely ($<1\%$ and $10\%$ error for various real control systems). 
Several potential applications 
are proposed to show that \mySimulatorName~could benefit the future design of NISQ compiler, architecture, etc. 

\end{abstract}

\section{Introduction}

Quantum Computing~(QC) has attracted great interest from both academia and industry during the last decades due to its strong potential in accelerating various important applications, e.g., integer factorization~\cite{shor1999polynomial}, database search~\cite{grover1996fast}, molecule simulation~\cite{peruzzo2014variational}. 
The second quantum revolution, \textit{transition from quantum theory to quantum engineering}~\cite{dowling2003quantum}, is leading us towards Noisy Intermediate-Scale Quantum (NISQ) era~\cite{preskill2018quantum}, when QC devices have fewer than 1000 qubits and are not large enough to support Quantum Error Correction~(QEC).
To make good use of such NISQ devices which suffer from limited qubit lifetime and imperfect operations,
more attention is 
given to NISQ system design and optimization in recent years, ranging across NISQ compiler~\cite{zulehner2018efficient, tannu2018case, siraichi2018qubit}, quantum control hardware architecture~\cite{fu2016heterogeneous, fu2017experimental, fu2018eqasm}, NISQ device~\cite{Google72Q,Intel49Q, IBM50Q, linke2017experimental}, etc.

Ideally, all these innovations should be evaluated on realistic NISQ hardware.
However, NISQ devices require extreme execution environment and most of them still remain in physics laboratories.
Existing QC cloud services, e.g. IBM Quantum Experience~\cite{IBMdevice}, Rigetti's QPU~\cite{RigettiQPU}, only provide limited access which can not satisfy the ever-increasing demand for experiments for evaluating new NISQ system designs.
These restrictions are blocking more researchers from getting into this area. 

Simulation can be a potential solution to this problem as NISQ system innovations can be proposed and evaluated without accessing realistic hardware.
Since a complete NISQ system consists of two major components, the quantum processor and its classical control system, a simulator for NISQ systems needs to meet the following requirements:

\begin{enumerate}

    \item \textbf{Simulating a Noisy Quantum Processor.} Quantum processors in NISQ era suffer from various noise effects. A simulator needs to be able to model the noises on realistic NISQ devices. The simulated output fidelity can help guide future NISQ system design. 
    
    \item \textbf{Simulating a Classical Control System.} The design of a control system can significantly affect the overall NISQ system performance, e.g., execution time. Such effects also influence the performance of the quantum processor. For example, longer execution time may exceed the qubit lifetime, which limits the size of a QC program that can be executed.
    
    
\end{enumerate}

Unfortunately, such a simulator that can satisfy these requirements is still missing.
Traditional architectural simulators, e.g., GEM5 \cite{power2015gem5}, GPGPU-Sim~\cite{bakhoda2009analyzing}, are designed for classical digital computing without the ability to simulate QC.
Existing QC simulators~\cite{chen2018classical, markov2008simulating, aaronson2004improved, anders2006fast, zulehner2018advanced, viamontes2004high, khammassi2017qx, smelyanskiy2016qhipster, wecker2014liqui, steiger2018projectq, quantumsim, jones2018quest, IBMqiskit}, no matter with or without noise effect considered, do not take the classical control system into consideration, lacking comprehensive modeling of a NISQ computing system.

\begin{figure*}[ht]
\centering
\includegraphics[width=2.0\columnwidth]{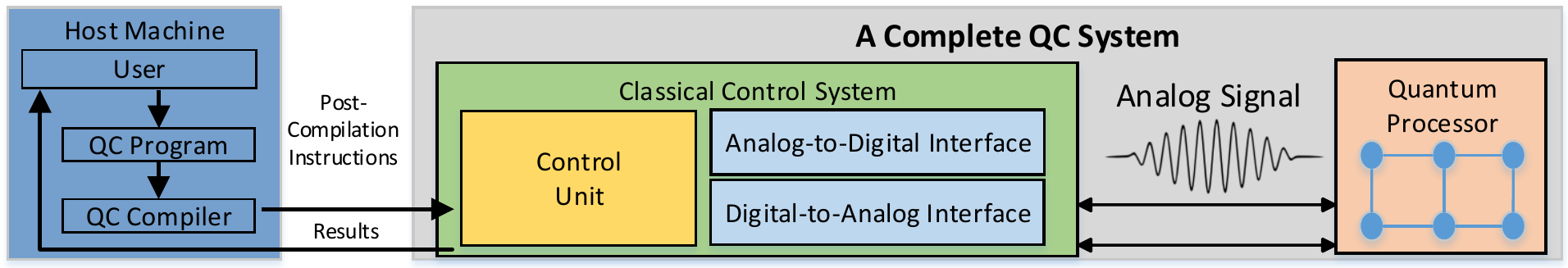}
\vspace{-10pt}
\caption{Schematic Overview of NISQ Computing System Architecture}
\vspace{-10pt}
\label{fig:overview}
\end{figure*}

In this paper, we propose a simulation framework, namely \mySimulatorName, for NISQ computing system design and evaluation.
\mySimulatorName~consists of one noisy QC simulator for the quantum processor, and one architectural simulation infrastructure to construct behavior models for the classical control system, leading to a comprehensive evaluation of NISQ systems and preparing for future design innovations.
The noisy QC simulator in \mySimulatorName~can adopt realistic error models and is accelerated by optimized Monte Carlo simulation. 
The architectural simulation infrastructure is specially designed for control system design.
Users can construct a behavior model for a classical control system with provided common hardware modules or with customized newly designed hardware components.
\mySimulatorName~currently focuses on the execution fidelity and timing simulation, which are both critical in NISQ system evaluation, while it is extensible to accommodate more simulation, e.g., power, reliability. 

To illustrate the design of \mySimulatorName, this paper demonstrates examples of how to adopt the noise model of a quantum processor and how to construct a control system based on the programming model and quantum processor interface requirement.
\mySimulatorName~is validated against real NISQ systems.
Several potential applications of \mySimulatorName~are proposed with examples to show that \mySimulatorName~can help with compiler and control system architectural design by providing early-stage evaluation and execution status analysis.
The main contributions of this paper can be summarized as follows:
\begin{itemize}


    \item We present a full system simulation infrastructure for comprehensive NISQ system modeling and early-stage evaluation.
 
 
    \item A noisy QC simulator is introduced with the ability to adopt the error model from realistic quantum processors. The proposed optimization can accelerate the error injection noisy simulation by $7\times$ on average without affecting the final results, compared with the brute-force simulation strategy on industrial simulator under selected benchmarks.
    
    
    \item We propose a simulation infrastructure for control system design.
    A mini control system is constructed by adopting realistic control hardware design.
    Several key hardware modules are provided, and they can also be reconfigured or customized by users.
    
    
    \item The entire simulation framework has been validated against realistic NISQ systems, including one superconducting quantum processor from IBM and two control systems from both IBM and Delft UT. The noisy simulator can capture the variance on the QC device and the control system simulator can simulate the timing behavior precisely ($<1\%$ and $10\%$ error for Delft and IBM's control systems, respectively).
    
    \item We propose three applications of \mySimulatorName, 1) full system performance evaluation, 2) design space exploration, and 3) finding new optimization opportunities, to show that \mySimulatorName~could benefit compiler optimization, control system design, etc.
    For example, our simulation suggests that increasing the number of Digital-to-Analog channels in the control system could provide at most 15\% execution time reduction.
    

\end{itemize}

The rest of this paper is organized as follows. We first introduce some background information about QC in Section~\ref{sec:background} and then provide an overview of \mySimulatorName~in Section~\ref{sec:overview}. The noisy QC simulator and its optimization are detailed in Section~\ref{sec:accelerate}. An example of constructing a mini control system in our simulation infrastructure is given in Section~\ref{sec:classicalcontrol}. We evaluate the QC simulation optimization and validate \mySimulatorName~in Section~\ref{sec:evaluation}. Several potential applications of \mySimulatorName~are proposed in Section~\ref{sec:casestudy}.
Some limitations and future works are given in Section~\ref{sec:futurework}. Related works are discussed in Section~\ref{sec:related}, and we finally conclude this paper in Section~\ref{sec:conclusion}.
\section{Background}\label{sec:background}

In this section, we will present a brief review of relevant background knowledge to help understand the NISQ computing system. 
We first introduce the fundamentals in QC, followed by an overview of NISQ systems.

\subsection{QC Basics}
\textbf{Qubit.}
Classical computing uses bits as the basic information unit with two deterministic states, `0' and `1', while QC employs qubits with basis states denoted as $\ket{0}$ and $\ket{1}$. 
The state of one qubit can be the linear combination of the two basis states, represented by $\ket{\Psi} = \alpha\ket{0} + \beta\ket{1}$, where $\alpha, \beta \in \mathbb{C}$ and $|\alpha|^2 + |\beta|^2 = 1$.
Two or more qubits can be in a superposition of more basis states.
For example, a two-qubit system can be in the state $\ket{\Psi} = \alpha_{00}\ket{00} + \alpha_{01}\ket{01} +  \alpha_{10}\ket{10} + \alpha_{11}\ket{11}$ and represented by a four-dimensional complex vector $(\alpha_{00},\alpha_{01},\alpha_{10},\alpha_{11})$.
In general, a $2^N$-dimensional vector is required to describe the state of a system with $N$ qubits.

\textbf{Quantum Operation.}
The state of a QC system can be manipulated by quantum operations.
The first type is quantum gates, which are unitary operators applied on one or more qubits to change the state vector.
The second type of operation is measurement operation, which will collapse the superposition state to the basis states with different probabilities based on the amplitudes in the state vector.

\begin{figure}[h]
\centering
\vspace{-5pt}
\includegraphics[width=0.75\columnwidth]{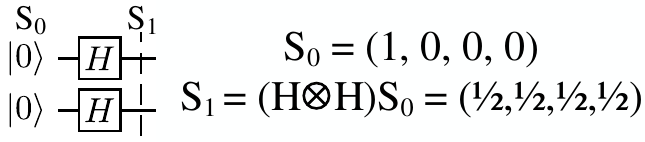}
\vspace{-10pt}
\caption{Example of Quantum Computation}
\vspace{-5pt}
\label{fig:qcexample}
\end{figure}

\textbf{Quantum Circuit and Computation.}
Quantum circuit is a diagram to represent a quantum program in the well-adopted quantum circuit model~\cite{nielsen2010quantum}.
Figure~\ref{fig:qcexample} shows an example of a quantum circuit and its computation.
On the left is the quantum circuit which contains two qubits and two $H$ gates (in the two squares).
The initial state is $S_0 = \ket{00}$ and its state vector $(1,0,0,0)$ is shown on the right.
To compute the state $S_1$, the two $H$ gates are applied on $S_0$ and the result is $S_1 = \frac{1}{2}\ket{00}+ \frac{1}{2}\ket{01}+ \frac{1}{2}\ket{10} +\frac{1}{2}\ket{11}$.
This process can be considered as a matrix-vector multiplication since the quantum system is linear and quantum gates are linear transformations.
The applied matrix is determined by the Kronecker product of the applied quantum gates.

\subsection{NISQ System}\label{sec:qcsystem}

Figure~\ref{fig:overview} shows a schematic NISQ computing system. 
On the left is a host machine, a classical computer which will interact with users and control the QC system. 
Users provide QC programs and the QC compilers will convert these programs to the basic instructions which can be executed by the control system.
The control system will further convert the instructions to control signals and send them to the quantum processor to implement different operations.

\textbf{Quantum Processor.}
The quantum processor is the core of the NISQ computing system, which can be implemented by different underlying technologies, e.g., superconducting quantum circuit~\cite{koch2007charge}, ion trap~\cite{lekitsch2017blueprint},  quantum dots~\cite{loss1998quantum}. 
The state of the qubits on the quantum processor is changed by external physical operations, e.g. micro-frequency electronic signals~\cite{chow2010quantum}, lasers~\cite{garcia2003speed}.
For the lack of QEC, the qubits are also affected by various noise effects~\cite{nielsen2010quantum}.
Unlike classical processors which work on digital signals, quantum processors are manipulated by analog signals.

\textbf{Classical Control System.}
A classical control system lies between the host machine and the quantum processor~\cite{van2018electronic}.
It converts post-compilation instructions into control pulse signals to control the quantum processor. 
The measurement results in analog form are also received from the quantum processor and converted to a digital form.
Such a classical control system provides a digital interface for the quantum processor and makes the NISQ system a co-processor of the host machine.

\section{Simulator Overview}\label{sec:overview}

\begin{figure}[h]
\centering
\includegraphics[width=1.0\columnwidth]{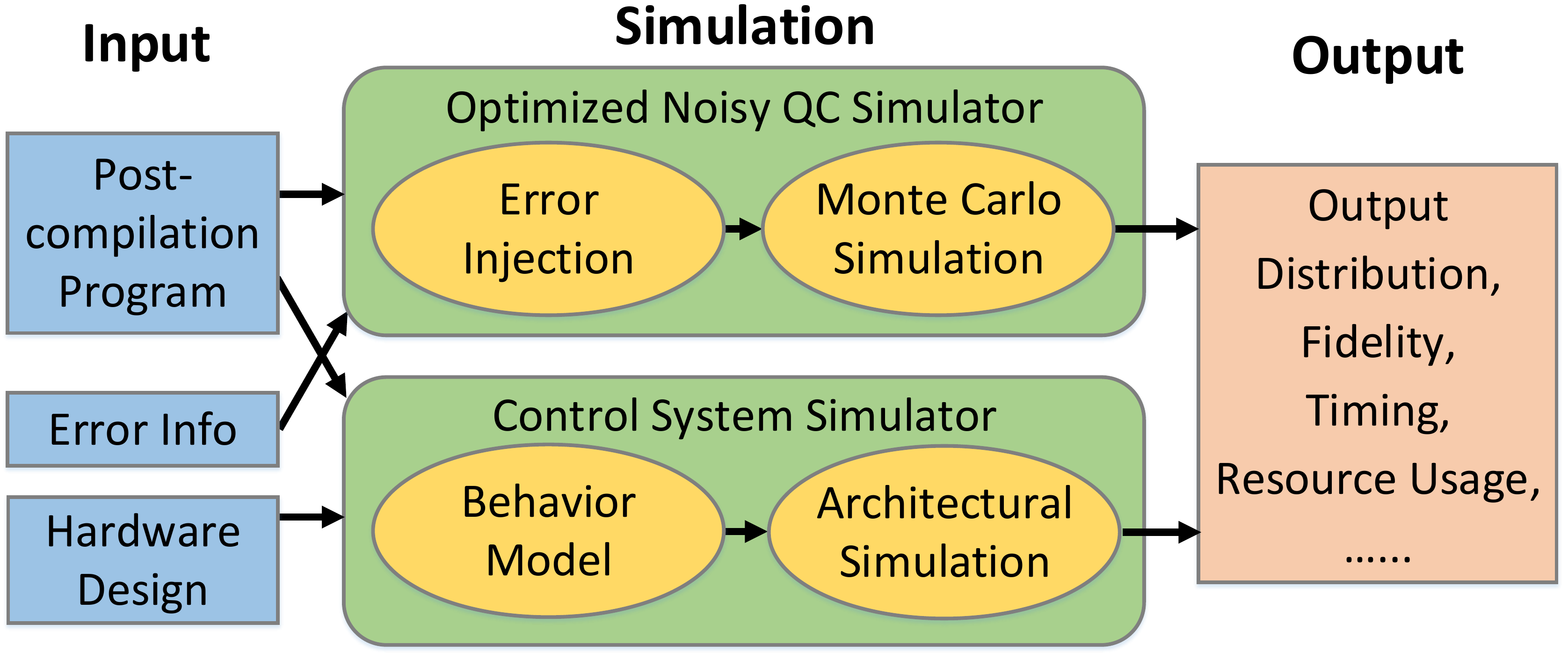}
\vspace{-15pt}
\caption{\mySimulatorName~Workflow}
\label{fig:workflow}
\end{figure}

In this section, we will provide an overview of \mySimulatorName, a simulation framework that contains a noisy QC simulator and a classical control system simulation infrastructure to cover the entire NISQ computing system. 
Both components are validated against realistic NISQ systems from IBM and Delft UT to guarantee the effectiveness of the simulation outcome. 
The workflow of \mySimulatorName~is illustrated in Figure~\ref{fig:workflow}.

\textbf{Input.}
The input required by \mySimulatorName~has three components, a post-compilation QC program, an error model, and a control system design. 
The instructions in the post-compilation QC program must be executable on the simulated hardware, which means all the quantum operations have been decomposed into hardware supported operations via compilation.
The rest two components are about the simulated NISQ system.
An error model should be provided to describe error operator, error position, and error probability on the simulated noisy quantum processor.
Users can define customized error model via the provided interface.
More accurate error model can come from the vendor or be characterized by physical experiments.
The hardware design of the control system is about the model of each hardware module in the simulated control system.
Users need to specify the output of each module under all possible input and how these modules are connected.
By default, \mySimulatorName~is pre-configured to be the baseline system model in the rest of this paper.
The input used in this paper for the baseline error model and control system design is provided for user reference. 

\textbf{Simulation.}
With the required input information, the two simulation components in \mySimulatorName~can provide comprehensive modeling of an entire NISQ system.
The noisy QC simulator uses the error information to construct an error model and generate error injection traces for the follow-up Monte Carlo~(MC) simulation.
The generated error injection traces will first be analyzed and reordered to eliminate redundant computation.
Then \mySimulatorName~will perform functional QC simulation for all the error injection traces and average the results, to obtain an output distribution and evaluate the fidelity.
On the other hand, the control system simulation infrastructure in \mySimulatorName~will use the provided hardware design to generate a behavior model for the simulated control system.
Traditional architectural simulation is then performed to model how the control system will execute each instruction of the input QC program and control the quantum processor.
Important information like the total execution time for a quantum program and the control hardware resource utilization rate can be simulated to evaluate the overall system performance.

\textbf{Output.}
The output from \mySimulatorName~will demonstrate key execution information of the simulated NISQ system.
The noisy QC simulator will provide the final output distribution in the MC simulation.
By comparing this result with error-free execution, \mySimulatorName~can evaluate the fidelity for one QC program execution on the simulated quantum processor.
The control system simulation will then provide detailed timing information for one execution.
More information like the occupation for each hardware component can also be collected to help locate the bottleneck in the simulated control system.



This section provides an overview of \mySimulatorName. In the next two sections, we will introduce the two simulation components in detail with examples of how \mySimulatorName~could simulate an existing NISQ system.
In Section~\ref{sec:accelerate}, we illustrate how to configure the error model based on IBM's public quantum processor information and how to accelerate the noisy QC simulation by eliminating redundant computations.
In Section~\ref{sec:classicalcontrol}, we will construct a mini control system in \mySimulatorName~based on real control systems.

\section{Noisy Simulation \& Optimization}\label{sec:accelerate}
In this section, we will illustrate how users can define an error model based on error information of a realistic device, followed by the optimization in our noisy simulator.
Quantum processors in the NISQ era are affected by noise effects.
A noisy QC simulator is designed and employed to capture the behavior of noisy quantum processors in \mySimulatorName. 
In general, simulation QC on a classical machine is a hard problem. 
Our noisy simulator can scale up to 20 qubits, which is the typical limit for standalone QC simulators~\cite{IBMqiskit,khammassi2017qx}.
However, noisy full-state QC simulation is still time-consuming since it requires simulating error-injected circuit many times to obtain an averaged result distribution.
We observe the computation redundancy inside such noisy simulation procedure and propose optimizations, achieving about $7\times$ speed-up on average without changing the output, compared with a brute-force noisy simulation strategy from industrial simulator~\cite{Rigettinoise}.

\subsection{Error Model Construction}~\label{sec:errormodel}
\begin{figure}[t]
\centering
\includegraphics[width=1.0\columnwidth]{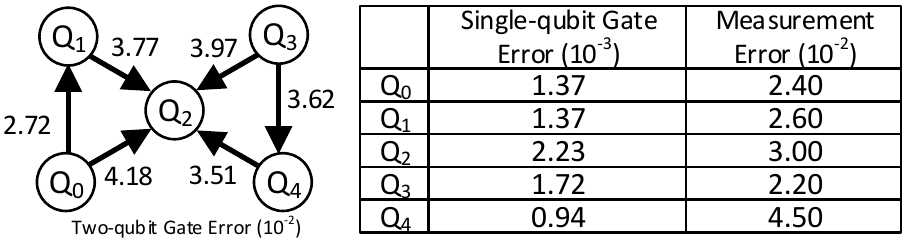}
\vspace{-20pt}
\caption{Error Rates on IBM Yorktown Chip~\cite{IBMdevice}}
\vspace{-10pt}
\label{fig:baselineqp}
\end{figure}

We use the error information from IBM's Yorktown 5-qubit superconducting quantum chip to construct an error model for the noisy QC simulation~\cite{IBMdevice}.
Figure~\ref{fig:baselineqp} shows the error rate data of IBM Yorktown chip.
Note that this information will change over time and we only sampled the results from one characterization.
On the left is the qubit coupling graph.
Each vertex represents a qubit and each edge in the graph means that a two-qubit Control-NOT (CX) gate can be applied between the two connected qubits.
The error rates for a CX gate applied on all edges are labeled.
On the right is a table showing the error rate of single-qubit gates and measurement operations for each qubit.
We will use this error rate data to construct an error model with error operator, error position, and error probability.

\subsubsection{Error Operator}\label{sec:erroperator}
Error operators are some special operators that will be randomly injected in the quantum circuit in order to model the noise effect in the QC program execution on noisy quantum hardware.
By default, \mySimulatorName~provides the basic Pauli error operators.
The three Pauli matrices, $X$, $Y$, and $Z$ (given in Equation~\ref{eq:pauli}), are basic error operators that can describe three types of errors, a bit flip error~($X$), a phase flip error~($Z$), and an error in which both a bit flip and a phase flip occur~($Y$).
\begin{equation}\label{eq:pauli}
X=\begin{bmatrix}
0 & 1 \\
1 & 0
\end{bmatrix},
Y=\begin{bmatrix}
0 & -i \\
i & 0
\end{bmatrix},
Z=\begin{bmatrix}
1 & 0 \\
0 & -1
\end{bmatrix}    
\end{equation}
Alternatively, \mySimulatorName~also supports customized error operators. Users can either define their customized set of error operators as a linear combination of these three operators, or directly specify their matrix representation.

The error information data from IBM do not specify the error operators.
As a result, we apply these three default Pauli matrices as the basic error operators.

\subsubsection{Error Position}
Error positions are the places where an error could possibly be injected in the simulated quantum circuit.
Errors can be injected after a gate or measurement if triggered by operations.
Some other errors like decaying from high-energy state $\ket{1}$ to low-energy state $\ket{0}$ or interacting with the environment can happen without an operation.
Such an error could appear at any place across the quantum circuit.

The error information from IBM is all about operation errors.
So that the constructed error model will only inject error operators after quantum gates or measurement operation.





\subsubsection{Error Probability}
After the error operators and positions are determined, we still need to know the probability for each error position with each error operator. 
Each time when we meet an error position during the simulation, we will randomly inject one error operator based on the error probability for each operator at this position.
IBM data have specified the error probability at each error position.
For example, an error will be injected after a single-qubit gate with a probability of $1.37\times 10^{-3}$ if this gate is applied on $Q_0$.

\begin{figure}[h]
\centering
\vspace{-10pt}
\includegraphics[width=1.0\columnwidth]{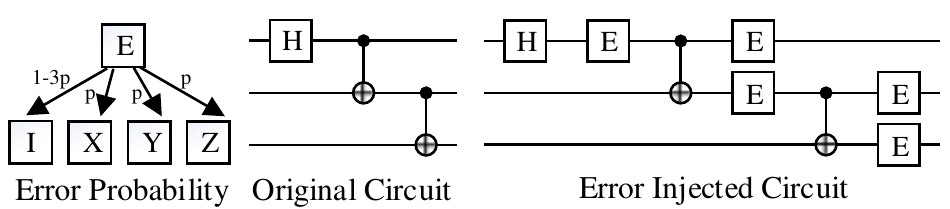}
\vspace{-20pt}
\caption{Depolarization Error Channel and Injection}
\vspace{-5pt}
\label{fig:depolarization}
\end{figure}

The error model construction has almost finished while a few more assumptions are still required.
First, the error probability is determined for each error occurrence but the error operator is not specified.
For this question, we assume the error comes from \textbf{depolarization error channel}, an important type of noise which converts a qubit to a completely mixed state with a small probability~\cite{nielsen2010quantum}. This error channel is widely used to model gate errors~\cite{IBMqiskit,khammassi2017qx} and suggested by IBM~\cite{IBMqiskit}. 
On the left of Figure~\ref{fig:depolarization} is the probability distribution of each error operator under depolarization channel.
An error operator $E$ can be one of $X$, $Y$, and $Z$ with the same probability $p$. 
Or it will become an identity operator $I$ with the probability $1-3p$, which means no error is injected in this position.

Second, for the measurement operation errors, since the error operator can only be applied to quantum states while the result after the measurement is a classical bit, we inject an error that flips the measurement result bit with the specified probability.
Finally, for two-qubit gates, we assume that error could happen on both qubits manipulated by the gate with equal probability independently.

\subsubsection{Error Injection}

To inject errors in the simulated circuit, error operator $E$ will be placed after each gate as shown in the error injected circuit in Figure~\ref{fig:depolarization}.
During the Monte Carlo~(MC) simulation of this error-injected circuit, every time when the QC simulator needs to simulate an error gate $E$, it will randomly replace $E$ by an error operator with the probability mentioned above or just ignore it without injecting an error.
Such MC simulation needs to run this circuit many times to generate an output distribution.
In the rest of this section, we will illustrate how to optimize this time-consuming simulation.

\begin{figure*}[t]
\centering
\includegraphics[width=2.0\columnwidth]{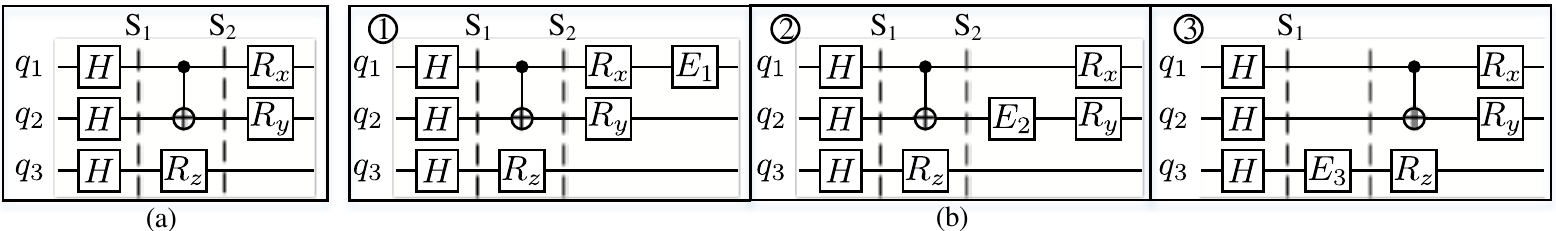}
\vspace{-10pt}
\caption{Motivating Example for Computation Redundancy and Execution Reordering}
\vspace{-10pt}
\label{fig:reorder}
\end{figure*}

\subsection{Noisy QC Simulator Optimization}

The proposed noisy QC simulator is optimized to exploit the computation redundancy among different error injection executions in the noisy QC simulation.
In this section, we will start from a motivating example to show the redundancy in the MC simulation.
Then we will show how to manage the executions to reduce the computation amount. 

\subsubsection{A Motivating Example}
Figure~\ref{fig:reorder} shows a motivating example to demonstrate the computation redundancy.
There are totally four error injection executions in this example, represented by four quantum circuits.
The first one in (a) is the original error-free execution. 
$S_1$ and $S_2$ are two intermediate states during the error-free execution.
The other three in (b) (labeled with \circledwhite{1}, \circledwhite{2}, and \circledwhite{3}) are error injected executions.
Each of them has one error operator occurred, represented by gate $E_{\{1,2,3\}}$.
To run the noisy QC simulation, all these four quantum circuits will be simulated and then averaged to obtain a distribution of the final output.
We can find that all the four quantum circuits are exactly the same before reaching $S_1$ state.
The state vector of $S_1$ is the same for all four execution since no errors are injected before $S_1$.
As a result, the computation from the initial state to $S_1$ can be shared by all four executions.
The state vector at $S_1$ only needs to be calculated and stored in one execution.
The rest three executions can start from the stored $S_1$ state instead of starting from the beginning.
Such redundancy exists at multiple locations across the error injection MC executions.
For example, the state vector at $S_2$ can be also be shared by the error-free execution and the first two error injected executions \circledwhite{1}\circledwhite{2}.


The motivating example above has shown computation redundancy among MC executions.
We can store some state vectors when we first reach such states and the results will be reused in the following executions.
However, the maximal number of state vectors we can store is limited since one state vector has $2^n$ amplitudes~($n$ is the number of qubits).
Although several techniques have been proposed to store the state vector in a compressed form~\cite{zulehner2018advanced,anders2006fast}, the memory requirement will still grow exponentially as the number of qubits increases.
To allow circuits with more intermediate states to be simulated efficiently, we introduce an execution reorder technique to reduce the number of concurrently maintained state vectors without loss of the benefit from the computation redundancy elimination.

\subsubsection{Execution Reorder}

Different execution order can significantly affect the number of states that need to be stored.
For the example in Figure~\ref{fig:reorder} (b), \circledwhite{1}\circledwhite{2}\circledwhite{3} is an inefficient MC execution order.
When running \circledwhite{1}, both the states $S_1$ and $S_2$ need to be stored so that \circledwhite{2} can start from $S_2$ and \circledwhite{3} can start from $S_1$.
An optimized execution order for this example can be \circledwhite{3}\circledwhite{2}\circledwhite{1}. 
When executing \circledwhite{3}, we only need to store state $S_1$.
The execution of \circledwhite{2} can directly start from the stored $S_1$ and then $S_1$ can be dropped since it is no longer used in the follow-up executions.
During the execution of \circledwhite{2}, $S_2$ will be stored and finally used when executing \circledwhite{1}. 
Consequently, only one state vector needs to be stored during the entire 
simulation process.
An optimized execution order reduced 50\% of memory requirement~(from two state vectors to one state vector) compared with a straight-forward order in this example.

In our noisy QC simulator, we first generate the MC execution traces without actually running the simulation.
The simulated quantum circuit is divided into layers, in which any two quantum operations are not applied on the same qubit. Error operators will only be injected at the end of each layer~(shown in Figure~\ref{fig:depolarization}). 
One execution trace will record the location and operator of each injected error.
These traces will be ordered by the location of the first injected error.
The traces with the first error injected in the first layer~(e.g., \circledwhite{3} in Figure~\ref{fig:reorder}) will appear at the beginning of the execution order, followed by those traces with the first error injected in the second layer~(e.g., \circledwhite{2} in Figure~\ref{fig:reorder}), and so on.

After the ordering procedure above, we begin our simulation by executing the first layer of the circuit with no error injected and store the state as $S_1$. 
This part of computation can be shared by all MC traces.
Then we will execute all the traces with errors first injected in the first layer.
If two or more error traces share the same first error~(injected on the same qubit with the same error operator), these traces will be grouped.
The simulation for these traces can be optimized recurrently if we consider $S_1$ as the initial state and let the remaining circuit after the first layer to be the simulated circuit.
After finishing the traces with first error in the first layer, we can execute one more layer without error and store the new state as $S_2$. 
Now $S_1$ can be dropped as no executions remaining will rely on it.
Additional memory space is only required when recurrent reordering happens because these traces with shared first error operator need the state vector after the shared error to help eliminate the computation redundancy among them.
The maximal number of state vectors we need to store is the recursion depth, which is small because the probability for two independent randomly generated traces to have $m$ shared error operators decreases exponentially as $m$ increases. 

The proposed noisy QC simulator will divide the input circuit into layers based on the error model and apply the optimized MC simulation. 
The final result is not changed since the simulated error-injected circuits are not changed and we just reuse computation across the MC simulation trials. 


\section{Control System Design}\label{sec:classicalcontrol}
Compared with conventional experiment instruments, integrated control system for QC is becoming more and more popular due to the demand for supporting larger scale QC systems~\cite{fu2017experimental}.
To build an architectural simulator for the QC control system, we investigated existing control systems designs~\cite{qin2017integrated, ryan2017hardware, salathe2018low, lin2018high, van2018electronic,fu2017experimental,Googlecontrol1,Rigetticontrol,IBMcontrol}.
The default control system model provided in \mySimulatorName, the mini control system, is constructed through abstracting the key components in real control systems.

In the rest of this section, we start from discussing the assumptions on the programming model, compiler, and quantum processor, because they will affect the interface of the control system.
Then we will introduce the hardware design and the behavior model of the mini control system.
This part and the noisy simulator in the last section make \mySimulatorName~a full NISQ system simulator.
Different from the full state QC simulator, this architectural control system simulator does not have a scalability issue.
If users do not need to simulate the output distribution, this part also can be used individually and simulate the control of a large-scale QC system.

\subsection{Assumptions}
Although programming and compilation should be done on the host machine and are not simulated in \mySimulatorName, some assumptions need to be made for them before we can continue to construct the architecture of a classical control system.
For the quantum processor, our assumption is only about the interface with the control system and does not affect the error models in the noisy simulation.

\subsubsection{Programming Model and Compiler}
\begin{figure}[h]
\centering
\includegraphics[width=1.0\columnwidth]{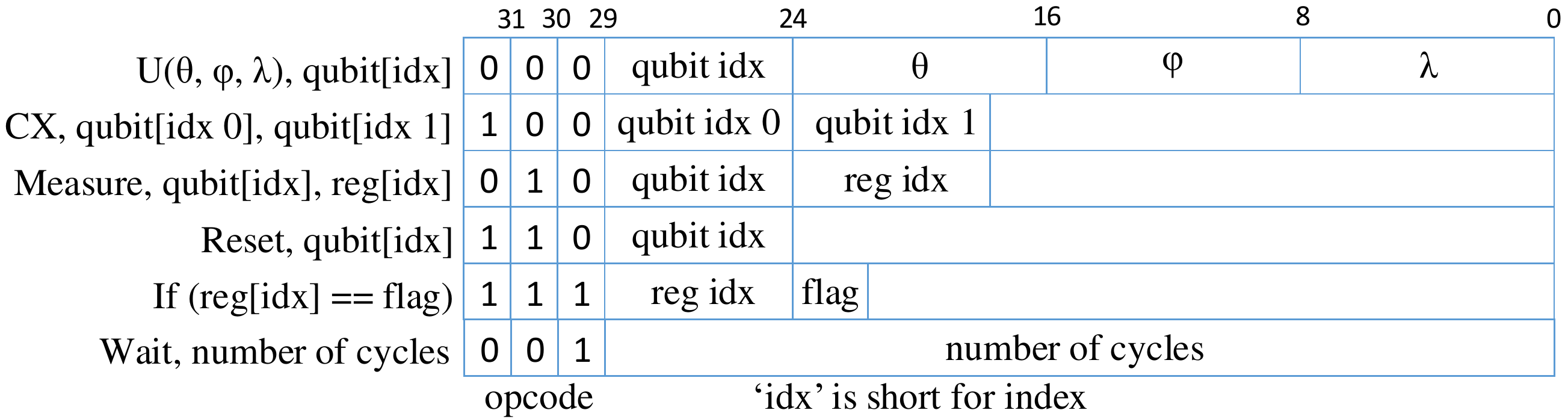}
\vspace{-20pt}
\caption{Instruction Encoding for Mini Control System}
\vspace{-5pt}
\label{fig:encoding}
\end{figure}

This mini control system accepts OpenQASM \cite{cross2017open}, the interface language of IBM's QC cloud service designed for small depth quantum circuits,  as the ISA.
OpenQASM is selected due to its rich benchmark resource and compiler support.
Quantum programs can be developed in high-level languages like Scaffold~\cite{abhari2012scaffold}, Quipper~\cite{green2013quipper}, or Q\#~\cite{svore2018q}, and then compiled to flattened OpenQASM format instructions.
However, some OpenQASM instructions are not executable so that we add some constraints for the program used in our mini control system.
There are only 5 types of instructions from OpenQASM remaining after compilation (the first 5 types in Figure~\ref{fig:encoding}). 
In addition, we add one 'Wait' instruction, which is critical in realistic control systems~\cite{fu2017experimental,IBMcontrol}, to enable more flexible timing control. 
Our control system will support this 6 types of instructions. 
For simplicity, the conditional instruction in the original OpenQASM standard is slightly modified and we only support one instruction in the branch based on one bit comparison result.
The quantum operations in the post-compilation instructions are in the Quantum ISA~(QISA) of the target quantum processor, which means the control signals for these operations are prepared and available.
A conditional instruction in OpenQASM is also included and will be managed inside the control system.
All the hardware constraints, e.g. the limited physical two-qubit gate availability, have been addressed during compilation optimization and the generated quantum program is completely hardware compatible.
All the post-compilation instructions have been pre-uploaded to an instruction memory in the control system. 
There is no communication between the host machine and the control system during the quantum program execution.

\subsubsection{Quantum Processor}
The quantum processor is assumed to be based on superconducting quantum circuit technology. 
The control signals for superconducting qubits are pre-calibrated micro-frequency electronic waveforms.
Adapting IBM's configuration~\cite{IBMbackend}, one single-qubit operation requires one control signal to be applied to that qubit.
One two-qubit gate needs three control signals applied to the two qubits and the resonator between them.
The quantum processor does not directly support operations on three or more qubits.

\subsection{Hardware Design}
With the assumptions above, users can specify the hardware design of the control system.
For each hardware module, users need to determine what internal states the hardware module should maintain, and the output under all possible inputs.
Moreover, users need to specify how the input and output ports of the hardware modules are connected in the hardware design.

\begin{figure}[t]
\centering
\includegraphics[width=1.0\columnwidth]{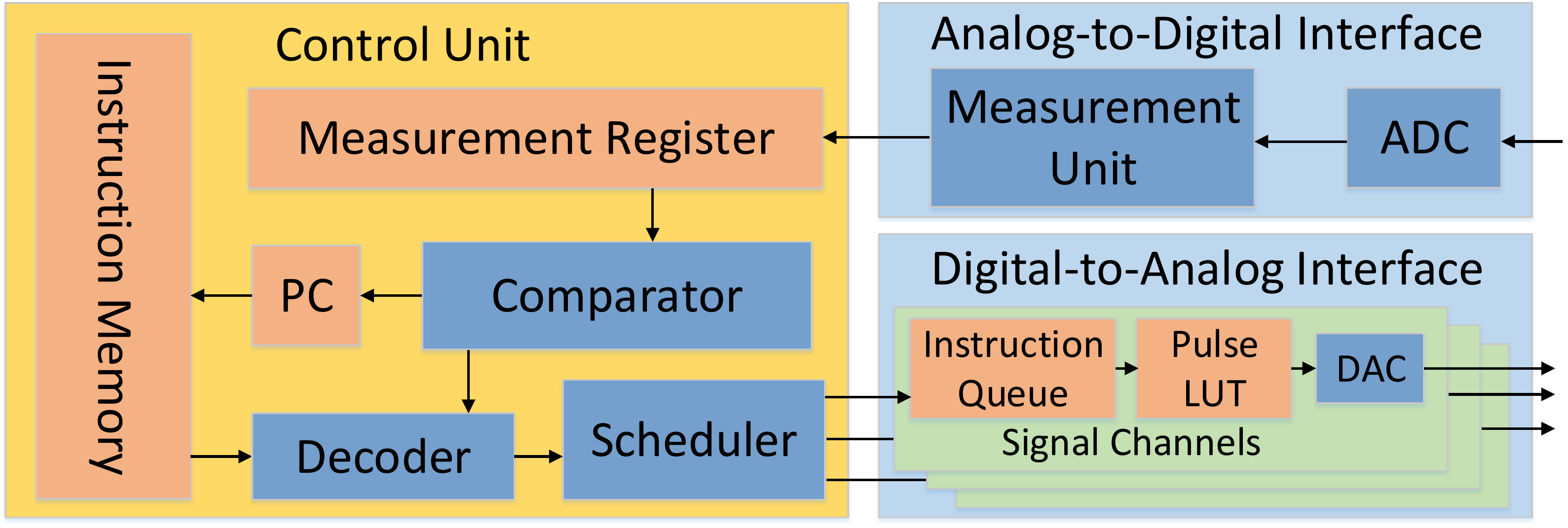}
\vspace{-20pt}
\caption{A Mini Control System}
\vspace{-10pt}
\label{fig:miniarch}
\end{figure}

As an example, a mini control system consisting of a control unit, a Digital-to-Analog~(DA) interface, and an Analog-to-Digital~(AD) interface, is shown in Figure~\ref{fig:miniarch}.
The hardware modules in the control unit are introduced as follows:
\begin{itemize}


    \item \textit{Instruction Memory.} This memory stores all the instructions. Since there is no existing binary encoding standard for OpenQASM~\cite{IBMopenqasm}. we assume that each instruction consumes 32 bits~(encoding shown in Figure~\ref{fig:encoding}). The input for this module is a memory address from Program Counter and the output is the instruction on that address which will be sent to a Decoder.
    
    \item \textit{Program Counter.} The Program Counter~(PC) records the address of the next instruction. It will automatically increase after one instruction is issued by the scheduler. It can also accept new address under conditional instructions.
    \item \textit{Measurement Register.} The measurement register stores the measurement results from the measurement unit. The comparator can read the measurement register.
    \item \textit{Decoder \& Comparator.} The decoder will decode those instructions in binary form fetched from the instruction memory. If it is a conditional instruction, the decoder will ask the comparator to read the measurement registers, do the comparison to determine the address of the next instruction. If an instruction needs to be applied on the quantum processor, the decoder will send the operation information to the scheduler.
    \item \textit{Scheduler.} The scheduler will decide which signal channel(s) will be used to apply an operation and send the operation to the instructions queue(s) of the signal channel(s). The operation dispatch policy is to find the signal channel(s) that can finish all the jobs in the queue(s) at the earliest time. The instructions are dispatched in order.
\end{itemize}

\textbf{Interface Design.}
The Mini Control System adopted the interface design from Quantum Control Box~(QCB)~\cite{fu2017experimental}, which is briefly introduced as follows.
For the DA interface, we employ three DA signal channels, the minimum requirement to implement two-qubit gates.
Each channel has an instruction queue as a temporary buffer for the instructions.
The waveform is implemented by a Pulse Look-Up-Table~(LUT), which can directly fetch stored pulse data, and we assume that all the pulse waveform data are already in the LUT.
A Digital-Analog-Converter (DAC) follows the Pulse LUT to generate analog signals.
For the AD interface, one AD channel will receive an analog signal from the quantum processor and convert it to digital form by an Analog-Digital-Converter (ADC).
The Measurement Unit will perform a weighted integration over the signal and then compare the results with a threshold value to determine whether the measurement result is 0 or 1.
All the channels in the AD/DA interface can connect to different qubits via switches.




\subsection{Behavior Model Generation}
After the hardware design is specified, \mySimulatorName~will generate a behavior model for the simulated control system.
A behavior model is about how the hardware will execute the given instructions.
In our mini control system example, only 5 types of basic instructions in OpenQASM standard~\cite{cross2017open} and the additional `Wait' instruction in Figure~\ref{fig:encoding} will appear after being compiled and flattened. 
The execution for these 6 types of instructions in the mini control system is listed here.
\begin{enumerate}
    \item \textit{$U(\theta,\phi,\lambda)$.} $U(\theta,\phi,\lambda)$ is a parameterized single-qubit gate. The decoder will send the instruction information to the scheduler and the scheduler will select one signal channel and put the instruction in the instruction queue. When this instruction is popped out, its control pulse will be fetched from the Pulse LUT, converted to an analog signal through DAC, and sent to the target qubit.
    
    \item \textit{CX.} CX is Control-NOT, the only supported two-qubit gate. Different from single-qubit gates, the scheduler needs to select three signal channels to complete this operation.
    
    
    \item \textit{Measure.} Measure is the measurement operation. The scheduler needs to choose one DA channel to send a special pulse and one AD channel will receive a feedback pulse. The Measurement Unit will determine the output and write the result to the Measurement Register.
    
    
    \item \textit{Reset.}  Reset is a single-qubit operation that reset the qubit to $\ket{0}$ state. In this mini control system, Reset is implemented by passive reset, which waits for $5\times T_1$ coherence time to let the qubit decay to $\ket{0}$ state.
    
    
    \item \textit{If.} This is a conditional instruction. The decoder will ask the comparator to read the measurement register, do the comparison to determine the address of the next instruction. If the condition is not satisfied, the next instruction will be ignored.

    
    \item \textit{Wait.} The control system will wait for a specific number of cycles before executing the next instruction.

    
   
\end{enumerate}

\subsection{Architectural Simulation}\label{sec:archanalysis}

After the behavior model is established, \mySimulatorName~will simulate the control system by executing the provided post-compilation instructions.
The post-compilation instructions are put into the \textit{Instruction Memory} first and PC is set to be the address of the first instruction.
Then, the configured NISQ control system will be simulated.

Besides simulating the execution time, \mySimulatorName~can also actively collect and record the states of all the hardware modules, e.g., the number of instructions in each instruction queue, the number of instructions executed, etc.
These statistical data can help locate the bottleneck in the system design. An example will be given in Section~\ref{sec:casestudy}.

\section{Evaluation}\label{sec:evaluation}
In this section, we first evaluate the speed-up and memory consumption of the optimized noisy simulation. 
Then we validate our simulator against real NISQ computing systems.

\subsection{Evaluating the Optimization in Noisy Simulation}

To evaluate the proposed optimization in the noisy simulation, we implemented a full state QC simulator with Python 3.4. The numerical noisy QC simulator is developed with Numpy 1.15.
All the experiments are executed on a server with Intel Xeon E5-2680 CPU. 
The operating system is CentOS 7.5 with kernel version 3.10.

\subsubsection{Experiment Configuration}
\textbf{Baseline.} The baseline noise simulation strategy is from Rigetti's QVM~\cite{Rigettinoise}, which repeats error injection simulation many times to generate an output distribution.

\textbf{Metric.} In order to perform a fair evaluation of our noisy simulator optimization, the metrics in this section are chosen to be independent of implementation and platform. 
For the computation time, we use the number of basic operations (matrix-vector multiplication) in full state QC simulation to indicate the computation amount.
For the memory consumption, we use the number of Maintained State Vectors~(MSVs) during the noisy simulation since the memory space for the state vectors, which will grow exponentially as the number qubits increases, dominates the memory consumption.


\color{red}{}
\color{black}

\textbf{Benchmarks.}
Table~\ref{tab:benchmark} shows the 12 quantum programs used in this experiment. They are collected from IBM OpenQASM benchmarks and prior work~\cite{IBMopenqasm, siraichi2018qubit}.
These benchmarks include Bernstein-Vazirani algorithm~(bv)~\cite{bernstein1997quantum}, Quantum Fourier Transform~(qft)~\cite{nielsen2010quantum}, Quantum Volume~(qv)~\cite{moll2018quantum}, Grover algorithm~\cite{grover1996fast}, Randomized Benchmarking~(rb)~\cite{knill2008randomized}, Modular Multiplication~(7x1mod15)~\cite{IBMqiskit}, and W-state~\cite{joo2003quantum}. 
The four columns on the right in Table~\ref{tab:benchmark} show the number of qubits and instructions in the post-compilation programs for each benchmark.
The selected programs have 5 or fewer qubits to be simulated on the IBM 5-qubit chip model (illustrated by Figure~\ref{fig:baselineqp}) and do not contain \textit{Reset} instructions.
The measurement instructions only appear at the end of each program so that there are no conditional instructions.
All the benchmarks only have $U(\theta,\phi,\lambda)$, $CX$, and $Measure$ instructions after compilation.

\begin{table}[h]
  \centering
  \caption{Benchmark Characteristics}
    \begin{tabular}{|c|c|c|c|c|}
    \hline
    Name~ & Qubit \# & U \#     & CX \#    & Measure \# \\
    \hline
    rb    & 2     & 9     & 2     & 2 \\
    \hline
    grover & 3     & 87    & 25    & 3 \\
    \hline
    wstate & 3     & 21    & 9     & 3 \\
    \hline
    7x1mod15 & 4     & 17    & 9     & 4 \\
    \hline
    bv4   & 4     & 8     & 3     & 3 \\
    \hline
    bv5   & 5     & 10    & 4     & 4 \\
    \hline
    qft4  & 4     & 42    & 15    & 4 \\
    \hline
    qft5  & 5     & 83    & 26    & 5 \\
    \hline
    qv\_n5d2 & 5     & 44    & 12    & 5 \\
    \hline
    qv\_n5d3 & 5     & 74    & 21    & 5 \\
    \hline
    qv\_n5d4 & 5     & 100   & 30    & 5 \\
    \hline
    qv\_n5d5 & 5     & 130   & 36    & 5 \\
    \hline
    \end{tabular}%
  \label{tab:benchmark}%
\end{table}%

\textbf{Compiler.} 
Quantum algorithms are usually developed for ideal device model while the allowed two-qubit gates are restricted by the available physical-qubit connections   on the hardware model~(shown in Figure~\ref{fig:baselineqp}).
Prior works have discussed how to overcome this problem by qubit allocation and remapping during compiler optimization~\cite{zulehner2018efficient,siraichi2018qubit}. 
In this paper, We choose the Enfield project~\cite{enfield}, which provides a dynamic programming based optimal qubit mapping in terms of gate count on IBM's 5-qubit devices~\cite{siraichi2018qubit}, as the compiler to generate hardware compatible quantum programs. 

\textbf{Experiment Method.} We run different numbers of trials (from 1024 to 8192) of the selected 12 benchmarks. 
The errors are injected based on the error model in Section~\ref{sec:accelerate}. 
Then we will compare the computation amount between the baseline and our optimized noisy simulation.
The effect on memory consumption of our reordering scheme is studied by comparing the number of MSVs with or without reordering the executions.


\subsubsection{Results}

\begin{figure}[h]
\centering
\vspace{-5pt}
\includegraphics[width=1.0\columnwidth]{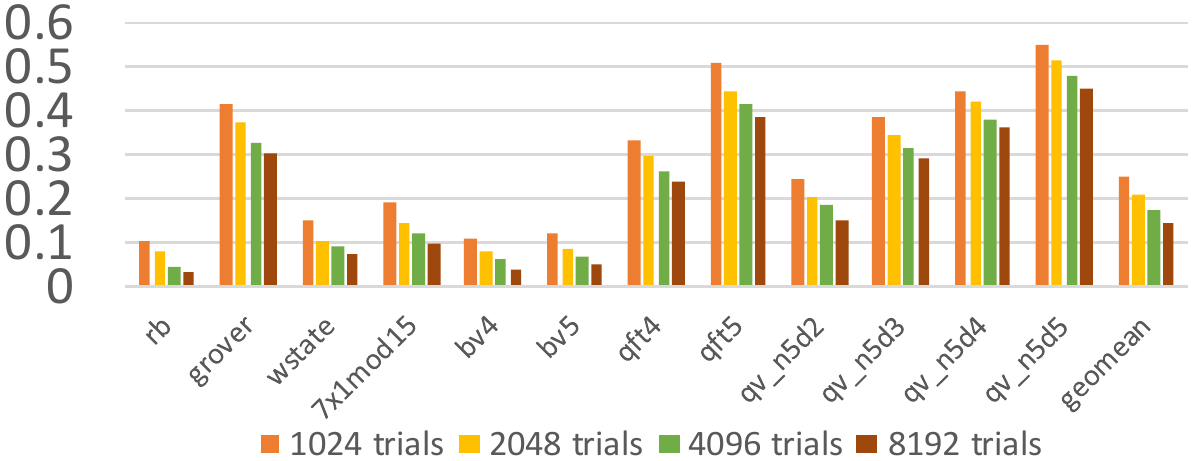}
\vspace{-20pt}
\caption{Normalized Computation Amount}
\label{fig:computationsaving}
\end{figure}

Our optimization modifies the simulation process to reduce runtime but does not affect the final output. 
Figure~\ref{fig:computationsaving} shows the computation saving for all benchmarks and different numbers of trials. 
The proposed optimization can save about 85\% of computation on average.
In the worst case when the benchmark is large (`qv\_n5d5'), the computation amount saving still achieves 57\% with 8192 trials. 
We can also find that the more trials we execute, the more computation we will save because more overlapped computation can be identified.

\begin{figure}[t]
\centering
\includegraphics[width=1.0\columnwidth]{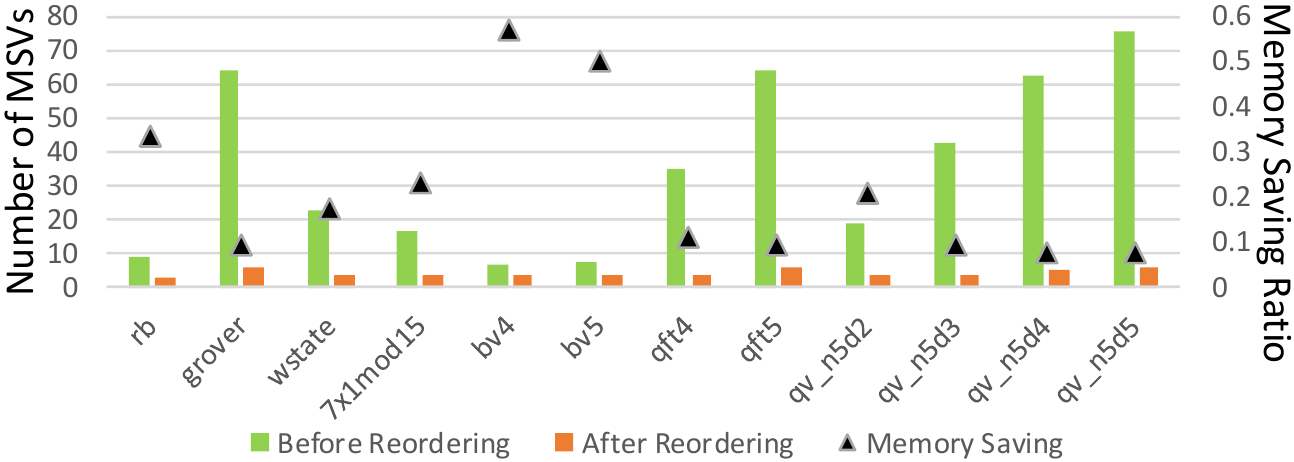}
\vspace{-20pt}
\caption{Memory Consumption for State Vectors. }
\vspace{-10pt}
\label{fig:memorysaving}
\end{figure}

Figure~\ref{fig:memorysaving} shows the memory saving in experiments with 1024 trials and this result does not significantly change when the number of trials increases from 1024 to 8192.
The number of MSVs is 3 for the smallest benchmark `rb' and only 6 in the largest benchmarks `qft5' and `qv\_n5d5'. As discussed in Section~\ref{sec:accelerate}, the number of MSVs will grow slowly since the probability for two trials to share the same $m$ injected errors decays exponentially with $m$.
As a result, the memory saving ratio will increase as the benchmark size increases.
For the small benchmark `bv4', about 43\% memory is saved. While for the largest benchmark `qv\_n5d5', our execution reordering technique can save about 92\% memory for MSVs.


\subsection{Simulator Validation}
In this section, both components in \mySimulatorName~are validated against realistic systems.
Due to the noise in the state-of-the-art NISQ systems, the output distribution of some benchmarks used in the last section will be hidden in noise and those benchmarks cannot be directly applied during validation experiments.
In the validation experiments, we carefully select validation methods, which will be explained later in this section, based on the capability of real QC systems.

\subsubsection{Noisy QC Simulator Validation}
\begin{figure*}[ht]
\centering
\includegraphics[width=2.0\columnwidth]{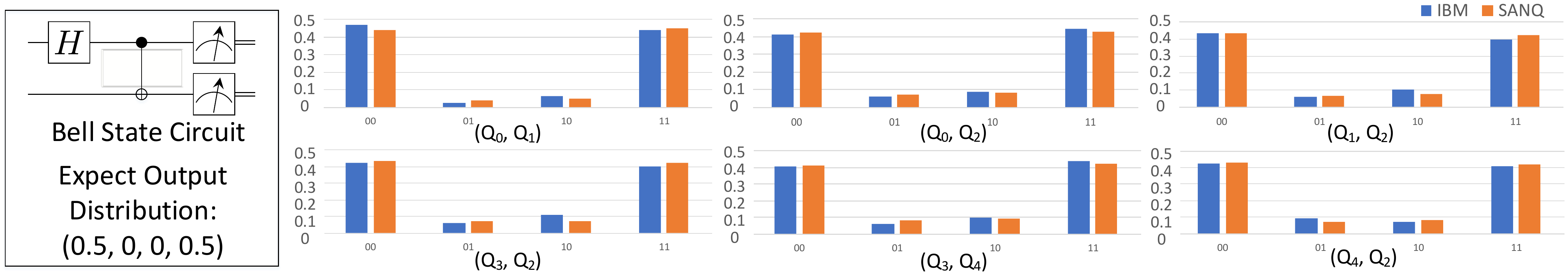}
\vspace{-10pt}
\caption{Simulation Results vs. Real Quantum Processor Execution}
\vspace{-10pt}
\label{fig:qpvalidation}
\end{figure*}
We validate the noisy simulator against IBM's Yorktown 5-qubit chip~(shown in Figure~\ref{fig:baselineqp}) and the error model is constructed in Section~\ref{sec:errormodel}.

\textbf{Validation Methodology.} Different from testing a classical digital device, a quantum processor has its unique benchmarking methodology. 
In experimental physics, Randomized Benchmarking~(RB)~\cite{knill2008randomized} is applied to each individual physical qubit and each connected physical qubit pairs. 
Such benchmarking method is widely accepted~\cite{kelly2014optimal,cross2018validating} and the error data in Figure~\ref{fig:baselineqp} is also from RB experiments run by IBM. 
Since our noisy simulator is targeting realistic devices, the validation experiments are designed based on the device calibration methodology.
We select the two-qubit Bell State program consisting of single-qubit gates, two-qubit gates, and measurement. We test it on all connected physical qubit pairs.
The experiment was repeated in 1024 trials and the output distributions are compared.


\textbf{Results.} 
Figure~\ref{fig:qpvalidation} 
shows the final output distributions of realistic execution results from IBM's real chip in blue, and the simulation results are shown in orange. 
The expected error-free output distribution in this experiment is $(0.5, 0, 0, 0.5)$, but the noise effect will make the output distribution slightly different.
For example, the output distribution for this experiment on $(Q_1,Q_2)$ qubit pair is about $(0.43, 0.06, 0.11, 0.40)$.
The simulation result for $(Q_1,Q_2)$ qubit pair is about $(0.42, 0.06, 0.07, 0.43)$, which is much close the realistic execution compared with the error-free result.
Among all the six experiments, the $(Q_0,Q_1)$ qubit pair has about 30\% lower error rate compared with other qubit pairs. 
The single-qubit error rate and measurement error rate on these two qubits are not significantly worse than others.
So the output of the experiment on $(Q_0,Q_1)$ on both the realistic quantum processor and our simulator show a distribution closer to the expected output compared with experiments on other qubit pairs.
Our simulator is able to capture the variation among different qubit connections by adopting the error rate information of a realistic quantum processor.


\subsubsection{Control System Simulator Validation}~\label{sec:controlvalidation}
\textbf{Validation Against Delft UT's Control System.} 
To validate our simulator against QCB~\cite{fu2017experimental},
The clock frequency is set to be $200MHz$.
Other key parameters are shown in Table~\ref{tab:baseline}.
The latency of single-qubit gates, two-qubit gates, and measurement operations are assumed to be $20ns$, $40ns$, and $300ns$, respectively. 

\begin{figure}[h]
\centering
\includegraphics[width=1.0\columnwidth]{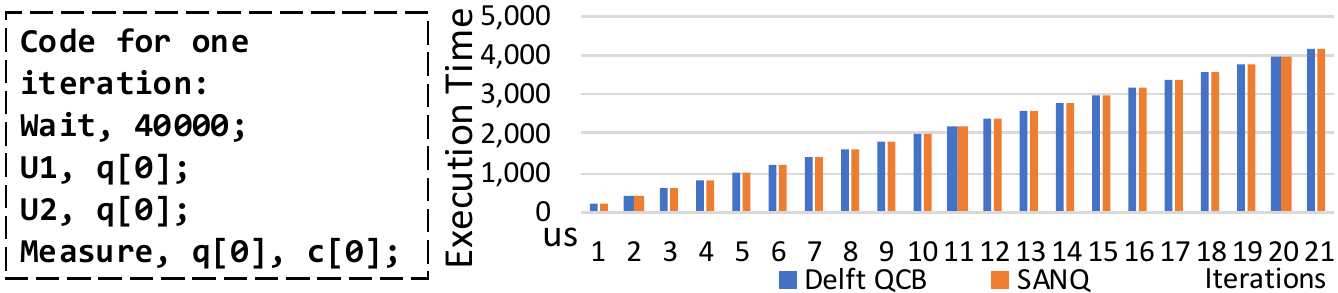}
\vspace{-20pt}
\caption{Validation Results For Delft QCB}
\vspace{-5pt}
\label{fig:validationdelft}
\end{figure}

To validate our simulator against QCB, we run the AllXY program\footnote{For details about AllXY program, please refer the QCB paper~\cite{fu2017experimental}.}, the original testing experiment for QCB~\cite{fu2017experimental}. The AllXY test program has 21 iterations and in each iteration, two single-qubit gates are applied to one qubit followed by a measurement operation. 
Figure~\ref{fig:validationdelft} shows the code for one iteration ($U1$ and $U2$ represent different single-qubit gates in different iterations) and the execution time on QCB and~\mySimulatorName. Our simulator could intimate the timing behavior of QCB with very low error ($< 1\%$) and the small error becomes negligible as the number of iterations increases. 

\textbf{Generality.}
Our simulation capability is not limited to QCB.
We compared the control system design for the superconducting quantum circuit from major vendors, including IBM~\cite{IBMcontrol}, Google~\cite{Googlecontrol1}, Rigetti~\cite{Rigetticontrol}, and Delft UT~\cite{fu2017experimental}. 
The control system architectures of them are similar, 
which suggests that our default design is representative. 
This architecture is also proved to be scalable and stable because Google is using a similar one~\cite{Googlecontrol1,Googlecontrol2} to control its 72-qubit chip by adding more hardware resources without changing the overall architecture.

\begin{table}[htbp]
  \centering
  \caption{Baseline Control System Model}
    \begin{tabular}{|c|c|c|}
    \hline
          & Single-qubit Gate & Two-qubit Gate \\
    \hline
    Latency & $20ns$  & $40ns$ \\
    \hline
    Channel & 1     & 3 \\
    \hline
    DA Channel \# & \multicolumn{2}{c|}{3} \\
    \hline
    AD Channel \# & \multicolumn{2}{c|}{1} \\
    \hline
    Measurement & \multicolumn{2}{c|}{Latency $300ns$, 1 AD Channel} \\
    & \multicolumn{2}{c|}{ and 1 DA Channel} \\
    \hline
    \end{tabular}%
    \vspace{-5pt}
  \label{tab:baseline}%
\end{table}%

\begin{figure}[h]
\centering
\vspace{-5pt}
\includegraphics[width=1.0\columnwidth]{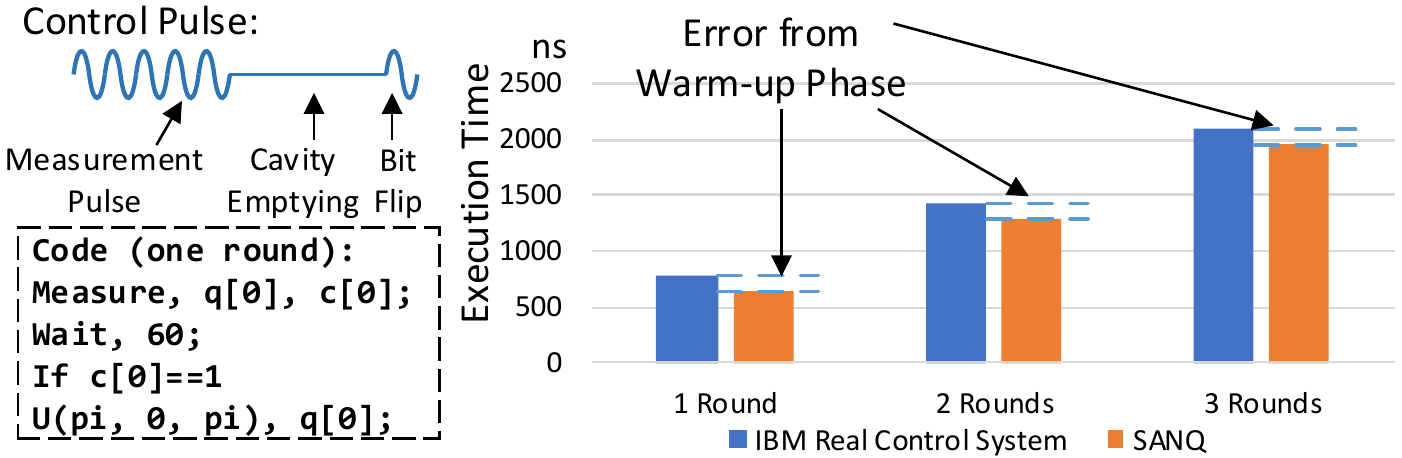}
\vspace{-20pt}
\caption{Validation Results For IBM}
\vspace{-10pt}
\label{fig:validationibm}
\end{figure}

\textbf{Validation Against IBM's Control System.} 
IBM's experimental control system model is different from the baseline.
The latency for single-qubit and two-qubit gates are $50ns$ and $300ns$, respectively, with 2 DA channels and 2 AD channels. 
The test program is Active Reset as shown in Figure~\ref{fig:validationibm} on the left. 
We first send measurement pulse to a qubit and then wait for 60 cycles for cavity emptying~(required by IBM's device). 
If the measurement result is $\ket{1}$, we apply a bit flip operation. 
This procedure is repeated for 3 times to guarantee a high reset fidelity.
The execution time of IBM's real control system and the simulation results are in Figure~\ref{fig:validationibm} on the right. 
The simulated execution time is close to that of IBM's real system. 
There exists a constant error (about $130ns$) which comes from the warm-up phase of the control system because such procedure before issuing the first instruction is not yet simulated in \mySimulatorName.
In summary, the average error ratio is 10\% and such error can be mitigated if we take the communication between the host machine and the control system into consideration, which will be addressed in our future work.

\color{black}




\section{Future Applications}\label{sec:casestudy}

In this section, we propose three future applications of \mySimulatorName~that are not available on existing QC simulators.
First, \mySimulatorName~can perform a comprehensive system performance evaluation by simulating both the quantum processor and the control system.
Second, \mySimulatorName~can perform design space exploration for the control system to guide future control hardware architecture design.
Third, by monitoring the utilization of the hardware components, \mySimulatorName~can help locate new optimization opportunities to improve NISQ system design. 
The rest of this section will provide three examples to illustrate the applications of \mySimulatorName~in detail.

\textbf{Baseline Configuration.}
The baseline quantum processor model in this section is from the IBM 5-qubit Chip~\cite{IBMdevice} and generated in Section~\ref{sec:accelerate}.
The control system model is the QCB validated in Section~\ref{sec:evaluation} with key parameters shown in Table~\ref{tab:baseline}. 
The baseline compiler remains the same with Section~\ref{sec:evaluation} and the benchmarks used are in Table~\ref{tab:benchmark}.

\subsection{System Performance Evaluation}\label{sec:sysperf}
We demonstrate the ability to perform a comprehensive system performance evaluation by comparing two different QC compiler optimization approaches on the qubit mapping problem.
One is the dynamic programming approach~(DYN) in Enfield, the baseline compiler~\cite{siraichi2018qubit}. The other one is a heuristic approach for efficient qubit mapping~(EFF)~\cite{zulehner2018efficient}.

\textbf{Experiment Design.}
We compile the 12 benchmarks with the two compilers mentioned above. 
Then we simulate the execution fidelity and time, from the noise QC simulator and the architectural control system simulator, respectively. 
Since some benchmarks are large and the correct output will be hidden by the noise on IBM's 5-qubit device~\cite{cross2018validating}, 
the term 'execution fidelity' used in this section is the ratio of error-free trial count over total trial count.
Both the quantum processor model and control system model used in compilation and simulation are the baseline models.

\begin{figure}[h]
\centering
\vspace{-5pt}
\includegraphics[width=1.0\columnwidth]{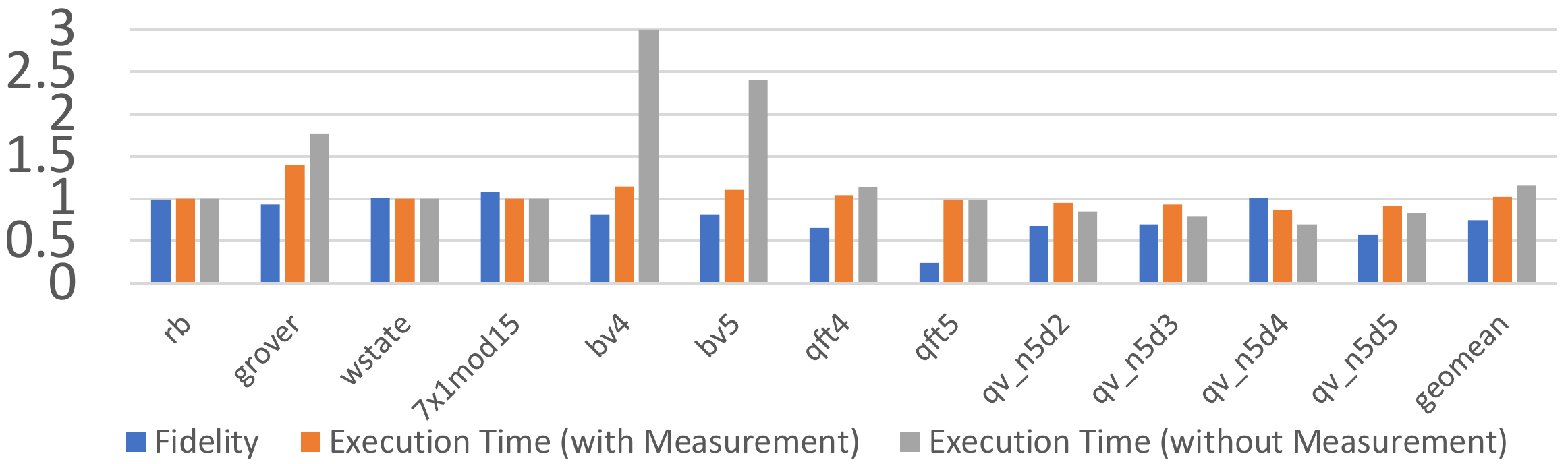}
\vspace{-20pt}
\caption{Normalized Simulation Result of EFF}
\vspace{-5pt}
\label{fig:compilercompare}
\end{figure}

\textbf{Results.}
Figure~\ref{fig:compilercompare}
shows that the execution fidelity and time (with and without measurement operations included) of EFF normalized to the results of DYN.
For two small benchmarks `rb' and `wstate', EFF and DYN generate the same code and the simulation results are the same for them.
In general, DYN is well optimized for CX gates and the execution fidelity is about 35\% better than that of EFF on average.
However, EFF also considered parallelism optimization.
For the `qv' benchmarks, the execution time is shorter for EFF even when the execution fidelity is still worse than EFF.
For the 'bv4' and 'bv5' benchmarks, they are small and the dominant factor in execution time is the CX gates so that EFF is much worse than DYN.
The original evaluation in the EFF and DYN papers~\cite{siraichi2018qubit,zulehner2018efficient} was based on the coarse-grained gate count and circuit depth metric in the generated program.
\mySimulatorName~generates consistent results to verify the optimality of DYN and the parallelism optimization in EFF. 
Moreover, \mySimulatorName~could perform fine-grained fidelity and execution time evaluation, preparing for deeper compiler optimization.

\begin{figure*}[t]
\centering
\includegraphics[width=2.0\columnwidth]{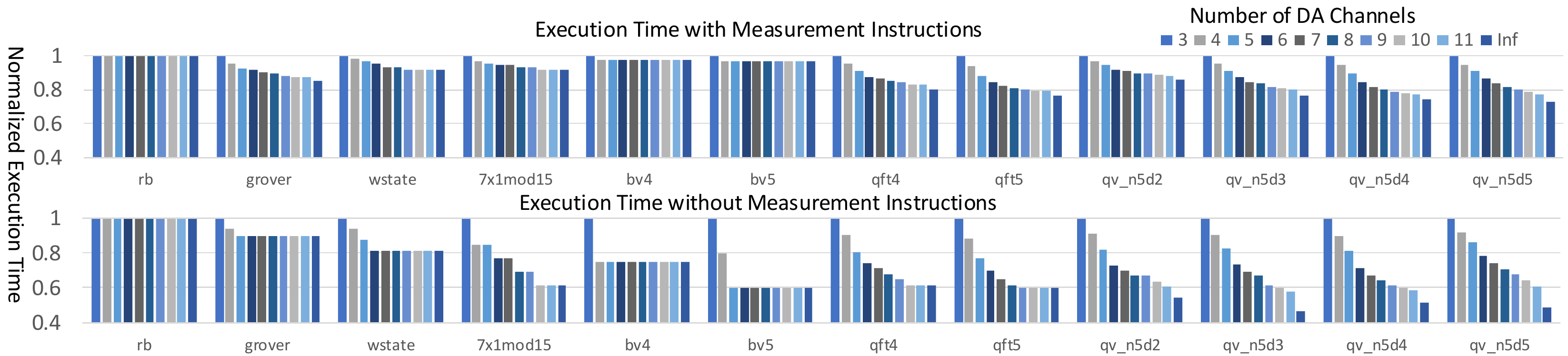}
\vspace{-10pt}
\caption{Execution Time Comparison with Various Numbers of DA Channels}
\vspace{-10pt}
\label{fig:channelresult}
\end{figure*}



\subsection{Design Space Exploration}
By simulating the classical control system, \mySimulatorName~is able to perform design space exploration to help guide the control system design.
This example focuses on the number of DA channels, which places an upper bound on the instruction parallelism.
For a quantum processor, instructions applied on different qubits can be executed in parallel theoretically.
However, the number of DA channels to send the control pulses is limited in a realistic control system.
The baseline employs three DA channels~(the same with QCB configuration~\cite{fu2017experimental}), which can support at most three simultaneous single-qubit operations or one two-qubit operation.
In this study, we investigate how the number of DA channels 
can affect the overall performance of a NISQ computing system.

\textbf{Experiment Design.} We vary the number of DA channels from three to eleven and simulate the execution time. 
All other configurations remain the same. 
In the end, we assume that there are infinite DA channels to remove this constraint.
This will show the ultimate limit if we continue to increase the number of DA channels.

\textbf{Results.} 
Figure~\ref{fig:channelresult} shows the execution time with various numbers of DA channels.
The results shown in the upper half include the measurement instructions.
All the benchmarks can benefit from more DA channels, except 'rb', which only has two qubits and is not constrained by the number of DA channels.
Larger size benchmarks can save more execution time than small size benchmarks.
When there are eleven DA channels, most benchmarks have been close to the upper bound with infinite DA channels, which is about 15\% on average since the execution is also limited by other effects, such as instruction dependencies.

Our simulation shows that the execution time of measurement instructions is the major limitation in this case study.
For all the experiments, the number of AD channels is always one which means that all the measurement instructions must be executed sequentially.
Moreover, the size of the selected benchmarks is small but the latency of measurement instruction is much longer than other operations in our quantum processor model~($300ns$ vs. $20\sim40ns$).
Fortunately, all the measurement operations are at the end of each benchmark and we can calculate the execution time before the measurement.
The execution comparison without the measurement instructions is provided in the lower half of Figure~\ref{fig:channelresult} and the average execution time-saving limit can achieve about 36\%.

\subsection{Finding New Optimization Opportunity}
\begin{figure}[h]
\centering
\includegraphics[width=1.0\columnwidth]{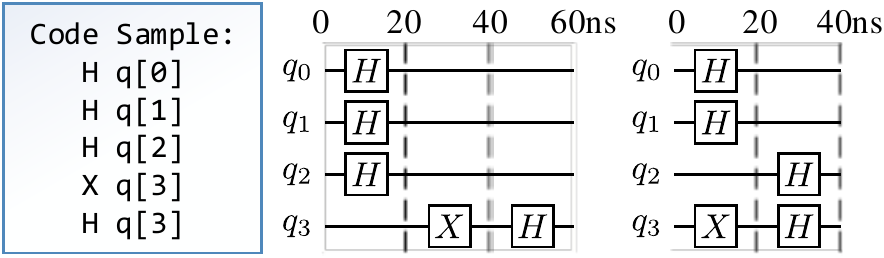}
\vspace{-15pt}
\caption{Example of Bottleneck}
\label{fig:bottleneck}
\end{figure}

The third example will show that \mySimulatorName~can suggest new optimization opportunities in NISQ system design by analyzing the execution status and locating the bottlenecks. 
For this example, we monitor the utilization rate of the DA channels in the system performance evaluation experiments~(in Section~\ref{sec:sysperf}).
Figure~\ref{fig:bottleneck} shows one bottleneck found in our experiments.
On the left are the first five instructions in `bv4' benchmark.
In this case, \mySimulatorName~finds that from $0ns$ to $20ns$, the number of instructions that is being executed is three and the DA channel utilization rate is 100\%.
But starting from $20ns$ to $60ns$, only one instruction is being executed and the utilization rate is just 33.3\%. 
The reason for this situation is discovered after looking into the execution details~(shown in the middle of Figure~\ref{fig:bottleneck}).
From $0ns$ to $20ns$, the first three instructions are 
executed in parallel.
The fourth instruction cannot be executed due to the DA channel constraint.
But from $20ns$ to $60ns$, only two instructions are executed because all the following instructions involve $q_3$ and cannot be executed before the fifth instruction.
As a result, the utilization rate of DA channels is only 33\% from $20ns$ to $60ns$. 



It is hard to locate such bottleneck through traditional program profiling or QC simulation without considering the actual control system.
\mySimulatorName~gives such opportunity to identify such hidden bottleneck, preparing for future system optimization.
For example, the bottleneck mentioned above can  potentially be solved in two ways.
One approach could be compiler optimization.
If the compiler knows that there are only three DA channels in the control system and hopes to reduce the execution time, a simple instruction reschedule can resolve this problem.
For example, Figure~\ref{fig:bottleneck} shows an example on the right.
The compiler can exchange the third and the fourth instruction without changing the circuit function.
The baseline control system can execute the $X$ gate on $q_3$ first. Then the remaining two $H$ gates can be executed in parallel.
Totally, the first five instructions now only consume $40ns$, saving 33\% of execution time compared with the original execution.
Another approach is to employ a more intelligent scheduling policy for the control system. 
The baseline considers one instruction at a time and only dispatch instructions in order.
This example suggests that a more powerful scheduler can consider more instructions ahead and issue instructions out-of-order to achieve a higher utilization of hardware resources. 

\section{Limitations and Future Work}\label{sec:futurework}

This paper provides a simulation framework for a whole NISQ system.
However, as an initial work in this area, \mySimulatorName~comes with some limitations.
In this section, we briefly discuss these limitations and our future plan.

\textbf{More Precise Noise Modeling.}
The proposed noisy QC simulator is equipped with widely used noise models.
However, errors in realistic hardware can be even more complex. 
For example, all errors are generated independently in our Monte Carlo simulation while error correlation actually exists and is being studied by physicists~\cite{aharonov2006fault, preskill2012sufficient}.
Deeper understanding of the mechanisms on the target QC platform will lead to more precise noise models.

\textbf{Advanced Quantum Control Architecture.}
The baseline control system is implemented with OpenQASM \cite{cross2017open}, a widely used intermediate representation for NISQ computing process.
This interface language is designed for small depth quantum circuit experiments on IBM's QC cloud service and lacks several important features for a control system ISA, e.g., efficient encoding, flexibility for quantum optimal control~\cite{werschnik2007quantum, leung2017speedup}.
For further research, \mySimulatorName~will adopt more advanced quantum control architectures, such as eQASM~\cite{fu2018eqasm}.

\textbf{Cooperating with Host Machine.}
\mySimulatorName~assumes that all the post-compilation instructions have been transferred to the control unit and does not include the communication between the host machine and the QC subsystem.
The assumption brings error in the simulation as discussed in Section~\ref{sec:controlvalidation}.
In the future, \mySimulatorName~can be integrated as a sub-module into an existing computer system simulator, e.g., GEM5~\cite{power2015gem5}, to include the modeling of the host machine and its cooperation with QC subsystem.

\section{Related Work}\label{sec:related}


\textbf{QC Simulator Optimization.}
Previous optimizations for QC simulators can be summarized into two categories.
Some simulators increase the simulation capability from algorithm-level~\cite{viamontes2009quantum,chen2018classical, markov2008simulating,aaronson2004improved, anders2006fast,zulehner2018advanced, viamontes2004high}. 
These works exploited sparsity or redundancy inside a single QC simulation process while the proposed optimization leverages the redundancy among multiple MC simulation executions.
The other type of optimizations is from computer system level, including vector instructions\cite{khammassi2017qx, smelyanskiy2016qhipster}, specialized linear algebra library~\cite{wecker2014liqui}, multi-thread~\cite{khammassi2017qx,smelyanskiy2016qhipster,steiger2018projectq}, distributed system~\cite{smelyanskiy2016qhipster, wecker2014liqui, steiger2018projectq}, GPU~\cite{quantumsim, jones2018quest}.
Our acceleration is from algorithm-level and is compatible with these system-level approaches.

\textbf{Noisy QC Simulator.}
Several existing QC simulators have supported error modeling and noisy simulation, e.g., IBM QISKit~\cite{IBMqiskit}, QX~\cite{khammassi2017qx}, Rigetti QVM~\cite{RigettiQPU}, `quantumsim'~\cite{o2017density,quantumsim}.
These simulators above can be used to model a realistic quantum processor while none of them is capable of evaluating an entire QC sub-system since the effect of classical control components is not considered.

\textbf{Classical Control System Design.} 
The electronic interface for quantum processors has been studied for small size cases~\cite{qin2017integrated, ryan2017hardware, salathe2018low, lin2018high, van2018electronic}.
Fu \textit{et al.} proposed QuMA, a microarchitecture, with accurate timing control, fast feedback control, etc. for a superconducting quantum processor~\cite{fu2017experimental}.
A cycle accurate microarchitectural-level simulator called QuMASim is developed for this specific architecture~\cite{fu2018quantum,zhang2018qumasim}. 
Leon \textit{et al.} also proposed a cycle accurate simulator for each hardware module in the control system without a complete microarchitecture~\cite{riesebos2017pauli}.
Dijk \textit{et al.} proposed SPINE, a toolset with a circuit simulator, for co-simulation of the electrical circuit and a spin-qubit-based quantum processor~\cite{van2018co}.
The proposed simulator combines both the quantum processor and the control system, providing fast and comprehensive NISQ modeling capability.


\section{Conclusion}\label{sec:conclusion}


This paper presents \mySimulatorName, a simulation framework for architecting NISQ computing systems. 
\mySimulatorName~consists of two components, an optimized noisy QC simulator and an architectural simulation infrastructure for the classical control system. 
The noisy QC simulator is equipped with flexible error modeling and optimized by computation redundancy elimination.
The architectural simulation infrastructure can construct behaviour models and evaluate control systems design decisions.
The usage of \mySimulatorName~is illustrated by adopting realistic error model and published control system design.
Three examples are given to show that \mySimulatorName~could benefit NISQ system design through comprehensive system evaluation and execution status analysis.
In conclusion, this paper proposes the first NISQ system simulator, allowing more researchers to participate in QC research and perform the early-stage evaluations for future innovations. 

\bibliographystyle{IEEETran}

\end{document}